\def\lesssim{{_ <\atop{^\sim}}}
\def\grtsim{{_ >\atop{^\sim}}}
\def\lax{{_ <\atop{^\sim}}}
\def\grtsim{{_ >\atop{^\sim}}}
\def\hkpc{h^{-1}{\ }{\rm kpc}}
\def\hMpc{h^{-1}{\ }{\rm Mpc}}
\def\hMsun{h^{-1}{\ }{\rm M_{\odot}}}

\def\Mvir   {M_{vir}}
\def\Rvir   {R_{vir}}
\def\Msun   {h^{-1} M_{\odot}}
\def\LCDM{{$\Lambda$CDM}}
\def\mxi{M_{X_i}}
\def\fIxi{f_{X_i}^{I}}
\def\yIxi{y_{I}(X_i)}
\def\yIIxi{y_{II}(X_i)}
\def\esng{E_{SN}^{g}}
\def\esn{E_{SN}}

\documentstyle[epsfig,mncite]{mn}
\begin{document}

   \title{On the supernovae heating of intergalactic medium}

   \author[Kravtsov \& Yepes]
          {
           Andrey V. Kravtsov$^{1,}$\footnotemark and
          Gustavo Yepes$^2$\\
          $^1$ Department of Astronomy, The Ohio State University, 
          140 West 18th Ave., Columbus, OH 43210-1173, USA\\
          $^2$Departamento de F\'{\i}sica Te\'{o}rica C-XI, Universidad Aut\'{o}noma de Madrid, 
              Cantoblanco 28049, Madrid, Spain
          }

   \date{Received ...; accepted ...}

   \maketitle

\begin{abstract}    
  We present estimates of the energy input from supernovae (SNe) into
  the intergalactic medium using (i) recent measurements of Si and Fe
  abundances in the intracluster medium (ICM) and (ii) self-consistent
  gasdynamical simulations that include processes of cooling, star
  formation, SNe feedback, and a multi-phase model of the interstellar
  medium. We estimate the energy input from observed abundances using
  two different assumptions: (i) spatial uniformity of metal
  abundances in the ICM and (ii) radial abundance gradients. We show
  that these two cases lead to energy input estimates which are
  different by an order of magnitude, highlighting a need for
  observational data on large-scale abundance gradients in clusters.
  Our analysis indicates that the SNe energy input can be important
  for heating of the {\em entire} ICM (providing energy of $\sim 1$
  keV per particle) only if the ICM abundances are uniform {\em and}\/
  the efficiency of gas heating by SN explosions is close to $100\%$
  ($\epsilon_{SN}\approx 1$, implying that all of the initial kinetic
  energy of the explosion goes into heating of the ICM).
  
  The SNe energy input estimate made using simulations of galaxy
  formation is consistent with the above results derived from observed
  abundances, provided large-scale radial abundance gradients exist in
  clusters.  For the cluster AWM7, in which such a gradient has been
  observed, the energy input estimated using observed metal abundances
  is $\sim 0.01$ and $\sim 0.1$ keV per particle for
  $\epsilon_{SN}=0.1$ and $\epsilon_{SN}=1$, respectively.  These
  estimates fall far short of the required energy injection of $\sim
  0.5-3$ keV per particle that appears to be needed to bring models of
  cluster formation into agreement with observations.  Therefore, our
  results indicate that, unless the most favorable conditions are met,
  SNe alone are unlikely to provide sufficient energy input and need
  to be supplemented or even substituted by some other heating
  process(es).

\end{abstract}


\section{Introduction}
\label{sec:intro}

\footnotetext{Hubble Fellow}
Hierarchical models of structure formation have been very successful
in explaining many observed properties of galaxies and galaxy
clusters.  Nevertheless, some puzzling problems remain open.  Several
theoretical studies have demonstrated that some heating of gas, in
addition to the heating during the gravitational collapse, is required
to explain the observed properties of the intracluster medium (ICM).
\scite{kaiser91:preheat} first showed that an early injection of
energy results in correlations and evolution of bulk cluster
properties (X-ray luminosity, gas temperature, etc.) that match
observations.  This conclusion was backed by numerical simulations:
\scite{evrard90:formation} was able to get a better fit to the X-ray
luminosity of the Coma cluster in his cosmological simulation by
preheating the gas to $10^7$ K ($\approx 0.9$ keV), while
\scite{nfw95} showed that pre-heated clusters matched the observed
slope of the correlation between X-ray luminosity and temperature.
Recent semi-analytical studies of cluster evolution have reached
similar conclusions
(\pcite{cavaliere97:lt,balogh99:preheat,valageassilk99:entropy,wu99:heating}).

The exact amount of required energy injection depends on the epoch,
and have been argued to be in the range of $0.5-3$ keV per gas
particle (\pcite{nfw95,cavaliere97:lt,balogh99:preheat,wu99:heating}).
Although the problem has been identified, it is not yet clear what
processes can provide the required heating. It is clear that
identification of these processes is crucial for a complete picture of
cluster formation.

The candidate process which has been discussed most is
supernovae-driven galactic winds. The gas of the galactic interstellar
medium (ISM) can be heated by supernovae explosions and acquire energy
comparable to or larger than its gravitational binding energy. This
heated gas can then flow away and result in additional heating if the
winds result in shocks when they encounter intergalactic gas. Although
there is observational evidence for such winds in present-day galaxies
(e.g., \pcite{heckman90:superwinds}), theoretical models of winds are
rather ill-constrained due to uncertainties in the efficiency of
conversion of supernovae (SNe) explosion energy into thermal energy of
the gas and other details.

Early estimates of possible energy input from SNe based on observed
metal abundances showed that SNe are plausible candidates (e.g.,
\pcite{white91:esn,david91:enrichment,loewenstein96}). However, in
these estimates it was assumed that the distribution of metals in the
ICM is uniform and that the efficiency with which energy of SNe
explosions can be converted into thermal energy of the gas is close to
$100\%$. Therefore, there is a need for detailed estimates using new
measurements of metal abundances in clusters and current galaxy
formation models which have become much more advanced and
sophisticated in the last several years.

Evaluations of SNe as a heating source have recently been performed 
by \scite{valageassilk99:entropy} and \scite{wu99:heating} using a
semi-analytical approach to galaxy modelling. These authors conclude 
that it is unlikely that supernovae are the only source of heating. 
\scite{valageassilk99:entropy} argue then that radiation from quasar 
population can provide the required heating much more easily. 

In this paper we repeat previous estimates of the possible SNe energy
input using updated values of observed ICM abundances and relaxing the
assumption of abundance uniformity, motivated by recent observations
(\pcite{ezawa97:awm7}; \pcite{finoguenov99}). We also make a separate
estimate for the cluster AWM7, for which the radial abundance gradient
was measured.  The details of this estimate are presented in
\S~\ref{sec:esnmetals}. We complement this analysis with direct
estimates of the energy input by counting the total number of
supernovae exploded in all cluster galaxies throughout their evolution
in self-consistent three-dimensional gasdynamical simulations of
galaxy formation. The simulations include cooling, star formation, SNe
feedback and a multi-phase model of the interstellar medium in
galaxies and have been shown to match many fundamental observed
correlations of galactic properties such as the galaxy luminosity
function, the Tully-Fisher relation and its scatter, the
color-magnitude sequence, and, perhaps most importantly, the evolution
of the global star formation rate in the Universe.  The details of the
simulations are described in \S~\ref{sec:esnsimul}.  The energy input
estimates are presented and compared in \S~\ref{sec:results} and
discussed in \S~\ref{sec:discussion}. We summarize our main results
and conclusions in \S~\ref{sec:conclusions}.

\section{Supernovae energy input from observed ICM metallicities}
\label{sec:esnmetals}

We will first estimate the energy input from SNe to the ICM using
observed metallicities of the cluster gas. We have based the estimate
on silicon (Si) and iron (Fe) abundances because these two elements
have been most accurately measured for a large sample of galaxy
clusters (\pcite{mushotzky96:abundances,fukazawa98:abundances}). We
use average ICM metallicities quoted in Table~\ref{table:yields} and
photospheric solar abundances of \scite{ag89:abund}
($n_{Fe}/n_H=4.68\times 10^{-5}$ and $n_{Si}/n_H=3.55\times 10^{-5}$).

Given that the mass of an element $X_i$, $\mxi$, within the cluster virial
radius is known, the number of SNe type I and II required to
produce this mass is equal to $\fIxi \mxi/\yIxi$ and $(1-\fIxi)
\mxi/\yIIxi$, respectively. Here $\fIxi$ is mass the fraction of the
element contributed by type I SNe, $\yIxi$ and $\yIIxi$ are the
mass-weighted yields of the element $X_i$ by SNe type I and
II respectively. The SNe energy input can then be obtained by multiplying the number of
SNe by the energy transferred to the gas during each SN explosion:
\begin{equation}
E_{SN}\approx \mxi\left(\fIxi\frac{\epsilon_{SNI}E_{SNI}}{\yIxi}+
(1-\fIxi)\frac{\epsilon_{SNII}E_{SNII}}{\yIIxi}\right),
\end{equation}
where we denote $E_{SNI}$ and $E_{SNII}$ energies released in
explosion of the two types of SNe, and $\epsilon_{SNI}$ and
$\epsilon_{SNII}$ are fractions of the released energy left after the
radiative losses during and after the explosion, which can be actually
transferred in the form of thermal and kinetic energy to the ambient
gas and lead subsequently to the increase of its entropy.  There is a
varying degree of uncertainty in our knowledge of the above
parameters.

First of all, our estimate of the mass of an element depends on the
assumption about uniformity of the observed metallicities. If strong
radial abundance gradients exist in clusters, the observed metallicity
is emission-weighted and therefore corresponds to the metallicity in the
cluster core.  Numerical simulations of cluster formation
(\pcite{metzler94,metzler97}) that include modelling of galaxy
feedback predict the existence of strong radial metallicity gradients in
the ICM, as well as patchy spatial distribution of metals. At present,
however, it is not clear whether large-scale abundance gradients are
universal in clusters. Although abundance gradients have been observed
in several clusters (see, e.g.,
\pcite{allenfabian99:metalgradients,dupke99:metalgradients} and
references therein), these are usually clusters that have a central cD
galaxy and exhibit signatures of a central cooling flow
(\pcite{allenfabian99:metalgradients}). The spatial extent of the
observed gradients coincides with that of the cooling flow region. It
is thus unclear whether such central gradients imply the existence of a
larger-scale gradient or they are simply due to the presence of a
central cD galaxy and cooling flow.

Currently, abundances in the fainter, outer parts of clusters can be
measured only for bright nearby systems. In a recent study,
\scite{ezawa97:awm7}, found strong large-scale metallicity gradients
in the nearby cluster AWM7. The observed iron abundance in this
cluster decreases from $\approx 0.5\pm 0.05$ solar within the central
$60\hkpc$ to $\approx 0.2\pm 0.2$ at $300-500\hkpc$.  The radially
averaged gradient can be well fitted by a $\beta$-model with a core
radius equal to that of the gas and $\beta=0.8$. The Ezawa et al.
measurement was the first in which the abundance gradient has been
found far beyond the cluster core radius. It is not yet clear how
common such large-scale gradients are. \scite{finoguenov99} show that
large-scale metallicity gradient are indeed observed in many clusters.
It is clear, however, that strong large-scale gradients are not
universal; for example, no strong gradient was detected in the Coma
cluster (\pcite{hughes93:coma}). Clusters may therefore exhibit a
variety of metal distributions and span a range in the ICM
metallicities. In support of this,
\scite{allenfabian99:metalgradients}, present evidence that clusters
without strong metallicity gradients have systematically lower metal
abundances.

For our purposes it suffices to consider two possible extreme
assumptions about the metal distribution. In reality the mass of
metals will likely lie in between the masses computed under these
assumptions. The first assumption is that the metallicity of the ICM is
spatially uniform.  Observationally, the metallicity derived from a
spatially unresolved spectrum is emission-weighted. It is clear then
that if a strong metallicity gradient is present in a cluster, the
total mass of metals may be significantly overestimated under the
assumption of spatial uniformity. Our second assumption is that the 
metallicity gradient of the form observed by Ezawa et al.:
$Z(r)=Z_0[1+(r/r_c)^2]^{-3\beta/2}$ is a universal property of the
cluster ICM. We will assume $r_c=100h^{-1} {\rm kpc}$ and $\beta=0.8$
(\pcite{ezawa97:awm7}) for all clusters, normalizing $Z_0$ to a value
such that $Z(r_c)$ is equal to the observed value of metallicity. Most of
the cluster emission comes from radii $<2 r_c$, providing an
approximate way to account for the emission-weighting of the
metallicity. The core radius of
$100\hkpc$ is larger than the best fit value for the cluster AWM7 but
is closer to a typical core radius of the gas distribution in rich
clusters. In addition to the estimate for the whole range of cluster 
masses, we will present the SN energy input for the specific case 
of AWM7 for which the abundance gradient has been observed and its
parameters measured. 

Another source of uncertainty is the relative importance
of type Ia SNe (parameter $\fIxi$) in the metal enrichment of the ICM 
(\pcite{loewenstein96,gibson97,nagataki98}). Therefore, in the
case of iron, we will treat $f_{Fe}^I$ as a free parameter and
calculate $E_{SN}$ for values ($\fIxi=0.0,0.5,1.0$). Silicon is a
special case, because $y_{II}(Si)\approx y_I(Si)$. This renders
$E_{SN}$ almost insensitive to a particular choice of $f_{Si}^I$. 
This insensitivity, together with the fact
that silicon abundance was fairly accurately measured by
\scite{mushotzky96:abundances} and \scite{fukazawa98:abundances},
effectively reduces the uncertainties and thus makes Si a very useful
element for our estimate.

The third major source of uncertainty is the yields predicted by different
theoretical models of SN explosions (\pcite{gibson97}). The yields of
SNe type II may depend on the initial metallicity of SNe, input
physics of a model, and other factors. In our estimate we will use
yields calculated by \scite{woosley:weaver95} for metal-poor
($Z/Z_{\odot}=10^{-4}$) SNe with explosion energy of $\approx
1.2\times 10^{51} {\rm ergs}$ (model A) and metal-rich
($Z/Z_{\odot}=1$) SNe with explosion energy of $\approx 1.2\times
10^{51} {\rm ergs}$ for SN of mass $\leq 25 {\rm M_{\odot}}$ and
$\approx 2\times 10^{51} {\rm ergs}$ for SN of mass $>25 {\rm
M_{\odot}}$ (Model B). Model B has a higher explosion energy for
very massive stars to reduce the effects of reimplosion of explosively
synthesized ejecta, thus increasing the yields. The yields for these
models approximately give the lower and upper limits of the current
theoretical predictions (see
\pcite{gibson97,nomoto97:yieldsSNII,nagataki98}).  Given the wide spread
in predictions of theoretical models, the supernovae energy input
estimates from the observed metallicities have the uncertainty of up
to $\sim 50\%$, in addition to other possible uncertainties.

We calculate the average yield of SNII by averaging the mass-dependent
yields with a stellar initial mass function (IMF):
\begin{equation}
\yIIxi=\frac{\int^{m_u}_{m_l}y_{II}(X_i,m)\phi(m)dm}
{\int^{m_u}_{m_l}\phi(m)dm}.
\end{equation}
For the IMF we assume the \cite{salpeter55} function, $\phi(m)\propto
m^{-2.35}$, with the lower mass limit of $m_l=12M_{\odot}$ and
$m_l=11M_{\odot}$ for the two yield models\footnote{We choose not to
use extrapolation and use mass limits of the yield grid of
\scite{woosley:weaver95}.  This does not significantly affect the
average yields.  \scite{gibson97} and \scite{loewenstein96} have used a somewhat larger
range of masses and obtained similar average yields.} and the upper mass limit of
$m_u=40M_{\odot}$. The SN with masses $<11M_{\odot}$ do not contribute
significantly to the metal enrichment, while stars of mass $>
40M_{\odot}$ are rare. Our choice of the Salpeter IMF does not affect
the average yields significantly: averaging with considerably
shallower ($\phi(m)\propto m^{-2}$) and steeper ($\phi(m)\propto
m^{-2.7}$) IMFs results in average yields that differ by less than
$10\%$ from the values for the Salpeter IMF. For  SN type Ia, the yields
appear to be independent of the SN mass. We use SNIa yields (see Table
1) predicted by the W7 model of \scite{nomoto97:yieldsSNIa}, which is a model 
of simple deflagration. The yields for the supernovae of type Ia and type II
used in our analysis are summarized in Table~\ref{table:yields}.

The energy released in a SN explosion ($E_{SNI}$ and $E_{SNII}$) also
depends on a variety of factors (e.g. the mass of supernova). However,
it can vary only by a factor of $\sim 2$ and we will therefore assume
for simplicity that $E_{SNI}=E_{SNII}= 1.2\times 10^{51} {\rm ergs}$
(e.g., \pcite{woosley:weaver86}). Not all of this initial kinetic
energy of explosion is retained by the ejected gas. Analytical
arguments (\pcite{larson74:sn,babulrees92:dwarfe}) and recent
numerical simulations (\pcite{thornton98}) suggest that at most $\sim
10\%$ of the initial kinetic energy of explosion can be ultimately
transferred to the ambient gas. In particular, \scite{thornton98} have
numerically studied the evolution of the ejected material for a
variety of densities and metallicities of the ambient gas and
concluded that regardless of the ambient density and metallicity,
$\grtsim 90\%$ of the initial energy acquired by ejecta during the
explosion is lost to radiation (see, however, discussion in
\S~\ref{sec:discussion}).  Therefore, we will assume that only $10\%$
of the explosion can actually be transferred to the ambient gas: i.e.,
$\epsilon_{SNI}=\epsilon_{SNII}=0.1$. Here, we neglect the dependence
of $\epsilon$ on environment, metallicity, and possible systematic
differences between $\epsilon_{SNI}$ and $\epsilon_{SNII}$. However,
according to \scite{thornton98}, radiation losses are $\sim 90\%$ or
more for most of the realistic environments and metallicities and the
value of $\epsilon=0.1$ is a reasonable upper limit for both types of
SNe. We note that the energy estimates from observed metallicities
presented in Figs.~\ref{esn}--\ref{esnpp} can be simply linearly
scaled up or down for other values of $\epsilon$.  Clearly, the energy
of SNe explosions is actually released into the interstellar medium
and needs then to be somehow transferred to the IGM. Even if such
transfer is possible, it is likely that it would result in additional
energy losses.  The estimates we make should therefore be considered
as the {\em upper} limits on the amount of the energy that could have
been available for the IGM heating.

All of the estimates are made assuming the low-density flat cold dark
matter model with cosmological constant (\LCDM). The contributions of
baryons, cold dark matter, and vacuum energy are: $\Omega_b=0.05$,
$\Omega_m=0.25$, $\Omega_{\Lambda}=0.7$, respectively.  We assume a
Hubble constant of $H_0=70{\ }{\rm km{\ }s^{-1}{\ }Mpc^{-1}}$.  The
cluster virial radius and mass for this model are defined at the
overdensity of $\approx 334$ and we assume the baryon fraction within
the virial radius is $f_b=\Omega_b/\Omega_m \approx 0.17$.

\begin{table}
\caption{Parameters adopted in the estimate of energy input from
   observed metallicities}
\label{param}
 \begin{center}
 \begin{tabular}{|lcccc} \hline
          &  Z    &  $y_I$ & $y_{II}^A$ & $y_{II}^B$   \\\hline
Si        & 0.65  &  0.158 & 0.124      & 0.158        \\
Fe        & 0.32  &  0.744 & 0.096      & 0.153        \\\hline
 \end{tabular}
 \end{center}
\label{table:yields}
 \end{table}

\section{Energy input from supernovae in numerical simulations of galaxy formation}
\label{sec:esnsimul}

   \begin{figure*}
   \resizebox{\hsize}{!}{\includegraphics{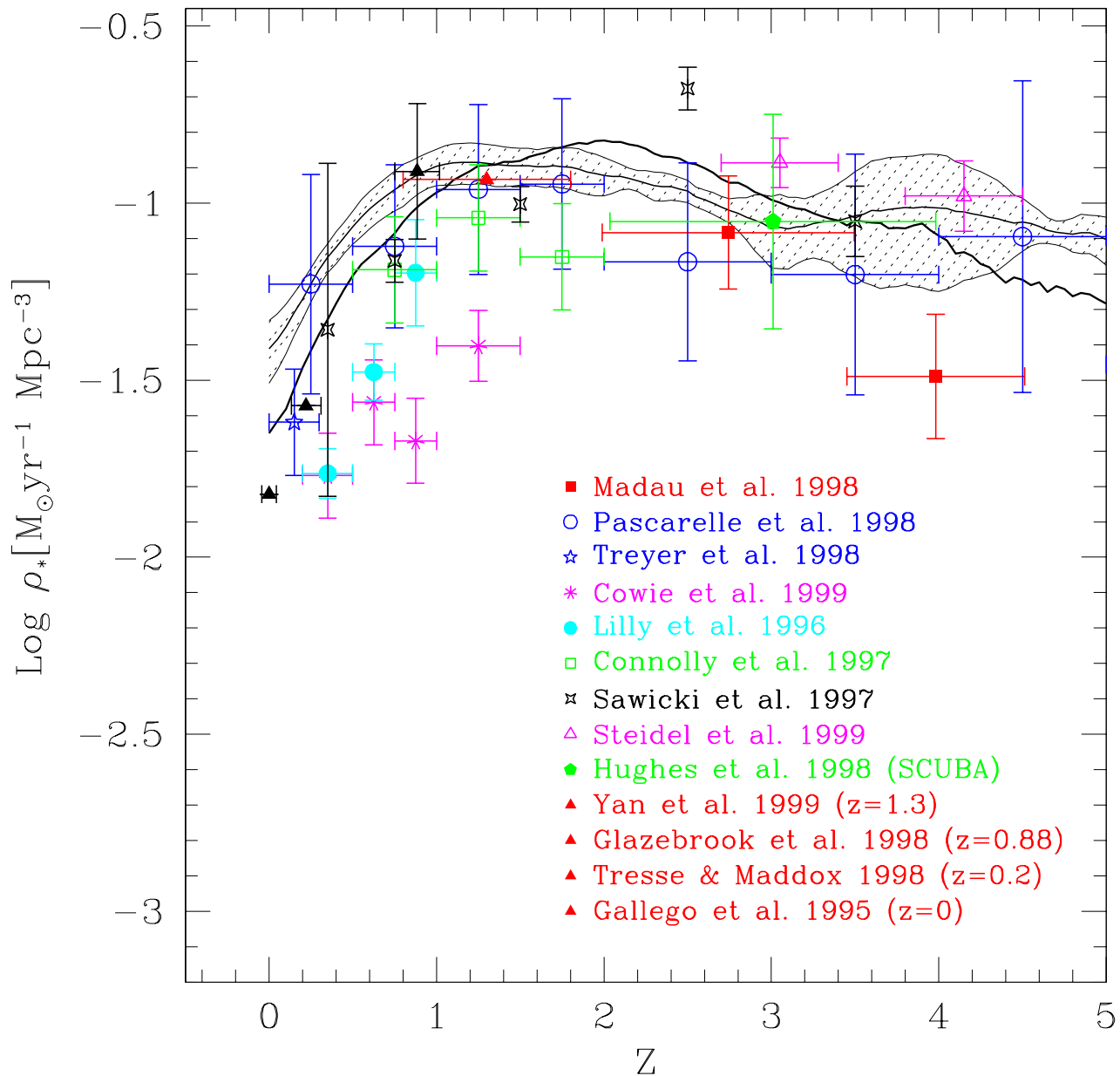}}
   \caption{Comparison of star formation history in the Universe
   (points), estimated from the UV and H$\alpha$ luminosity densities,
   corrected for dust extinction, with the star formation history
   obtained in the gasdynamical simulations used in this paper. The
   {\em solid lines} indicate the average and scatter of the star
   formation history in eleven $5$ Mpc runs, while the {\em thick
   solid line} shows the star formation rate in $12$ Mpc run.  All of
   the data points have been converted to the {\LCDM} cosmology with
   $\Omega_0=1-\Omega_{\Lambda}=0.35$ and the Hubble constant of
   $H_0=70\rm {\ }km{\ }s^{-1}{\ }Mpc^{-1}$.  }  \label{sfr}
   \end{figure*}

   \begin{figure*}
   \resizebox{\hsize}{!}{\includegraphics{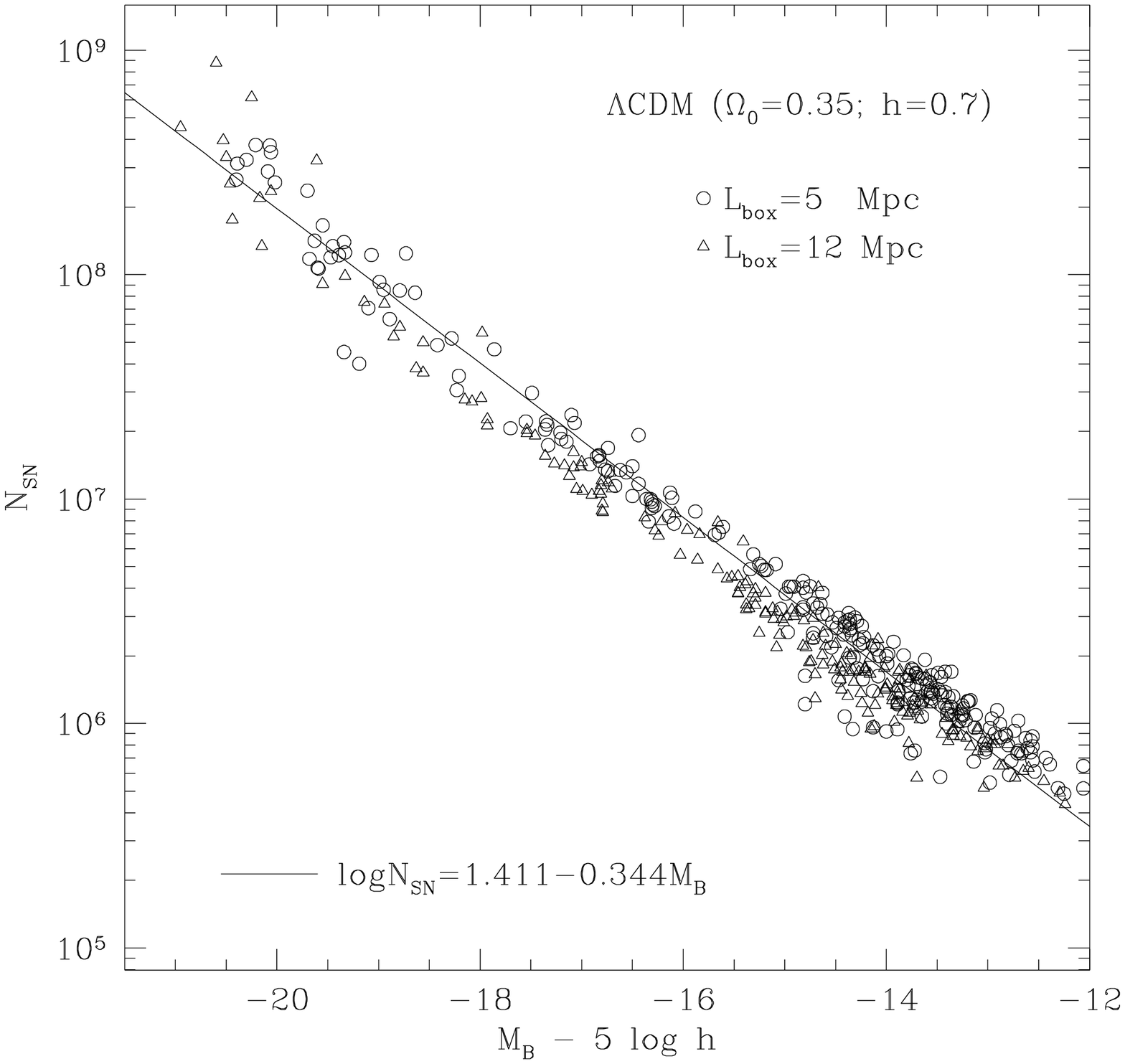}}
   \caption{Correlation of the $z=0$ $B$-band absolute magnitude of
   simulated galaxies and number of SNe type II exploded in the
   galaxy throughout its history. The {\em open circles} represent
   simulated galaxies in $5$ Mpc runs, {\em triangles} represent
   galaxies in the $12$ Mpc run, and the {\em solid line} shows a linear fit
   to the correlation. Number of SNe, $N_{SN}$, is estimated as
   $N_{SN}=0.12M_{\ast}/22 M_{\odot}$, where $M_{\ast}$ is the mass of gas
   converted into stars, $0.12$ is the fraction of the mass converted
   into stars with mass $> 10 M_{\odot}$, and $22 M_{\odot}$ is the
   IMF-weighted mean supernova mass. Both numbers assume a Salpeter IMF.
   }  \label{nsnm} \end{figure*}

The question we now ask is what energy, $\esng$, can be expected to
have been released in all the SNe in cluster galaxies throughout their
evolution? This question can be answered only in the framework of a
self-consistent model of galaxy formation. There are currently two
independently developing approaches to modelling galaxy formation and
evolution: semi-analytic models (SAMs; e.g.,
\pcite{kwg93:sam,bcf96:sam,sp98:sam} and references therein) and
numerical models (e.g., \pcite{katz92:galform,sm95:galform,yk3}). In
this section we will make an estimate of $\esng$ using numerical
simulations of galaxy formation. The numerical techniques and physical
ingredients of the model are described in \scite{yk3}. The model
includes a self-consistent treatment of the dark matter and baryonic
components and effects of cooling, star formation, and SNe feedback.
Simulations include a multi-phase model of interstellar medium. Note
that these simulations account only for SN type II so the contribution
from SNI is therefore neglected in the estimate of $\esng$.

The ideal simulation for our purpose would be a full modelling of
cluster formation that would include formation of the cluster
galaxies, their starformation and feedback. However, with the
numerical code used here, this would require a significant sacrifice
in the spatial dynamic range and mass resolution and would make it
impossible to follow reliably the starformation and feedback
processes. Such simulation awaits future higher dynamic range
simulations using adaptive mesh refinement technique.  We choose the
following compromise. We use many small-box galaxy formation
simulations to determine statistically the number of supernovae,
$N_{\rm SN}$, that is expected to explode in a galaxy of a given
absolute magnitude, $M_{\rm B}$.  We then use this relation and assume
a galaxy luminosity function in clusters to estimate how many
supernovae could have exploded in a cluster of a given mass. The
energy released in these SN explosions would provide an upper limit on
the amount of energy available for IGM heating. While the number of
SNe could be estimated by {\em assuming} a particular $N_{\rm SN}-M_{\rm B}$
relation, the use of simulations in this study spares us from making
this additional assumption.  As we describe below, the simulations
reproduce many of the observed galactic properties which provides 
support to the used $N_{\rm SN}-M_{\rm B}$ relation.

The simulations of the {\sl COBE}-normalized {\LCDM} model
($\Omega_0=1-\Omega_{\Lambda}=0.35$; $\Omega_b=0.026$; $H_0=70{\rm {\
}km{\ }s^{-1}{\ }Mpc^{-1}}$, where $\Omega_0$, $\Omega_b$, and $H_0$
are present day values of the matter and baryon densities and the
Hubble constant, respectively) used here are described in
\scite{elizondoetal99:apj}.  A total of 11
simulations were run from different realizations of initial
conditions. The size of the simulation boxes was fixed to
$L_{box}=3.5\hMpc=5 {\rm Mpc}$ and the simulations were run using
$128^3$ grid cells and particles which gives mass and spatial
resolution of $\approx 2\times 10^6\hMsun$ and $\approx 27\hkpc$,
respectively. A total of $240$ galaxies, $140$ of which have
$M_B(z=0)<-14$, were formed in all the runs combined.  The observed
color-magnitude diagram, luminosity function (LF), and Tully-Fisher
relation of low-redshift galaxies are reproduced well by the simulated
galaxies (\pcite{elizondoetal99:newa,elizondoetal99:apj}).
Simulations used here were done assuming SN feedback parameter of
$A=200$ (see \pcite{yk3}).  This value of the parameter means a
moderate efficiency of supernovae feedback and, correspondingly,
relatively high star formation rate.

   \begin{figure*}
   \resizebox{\hsize}{!}{\includegraphics{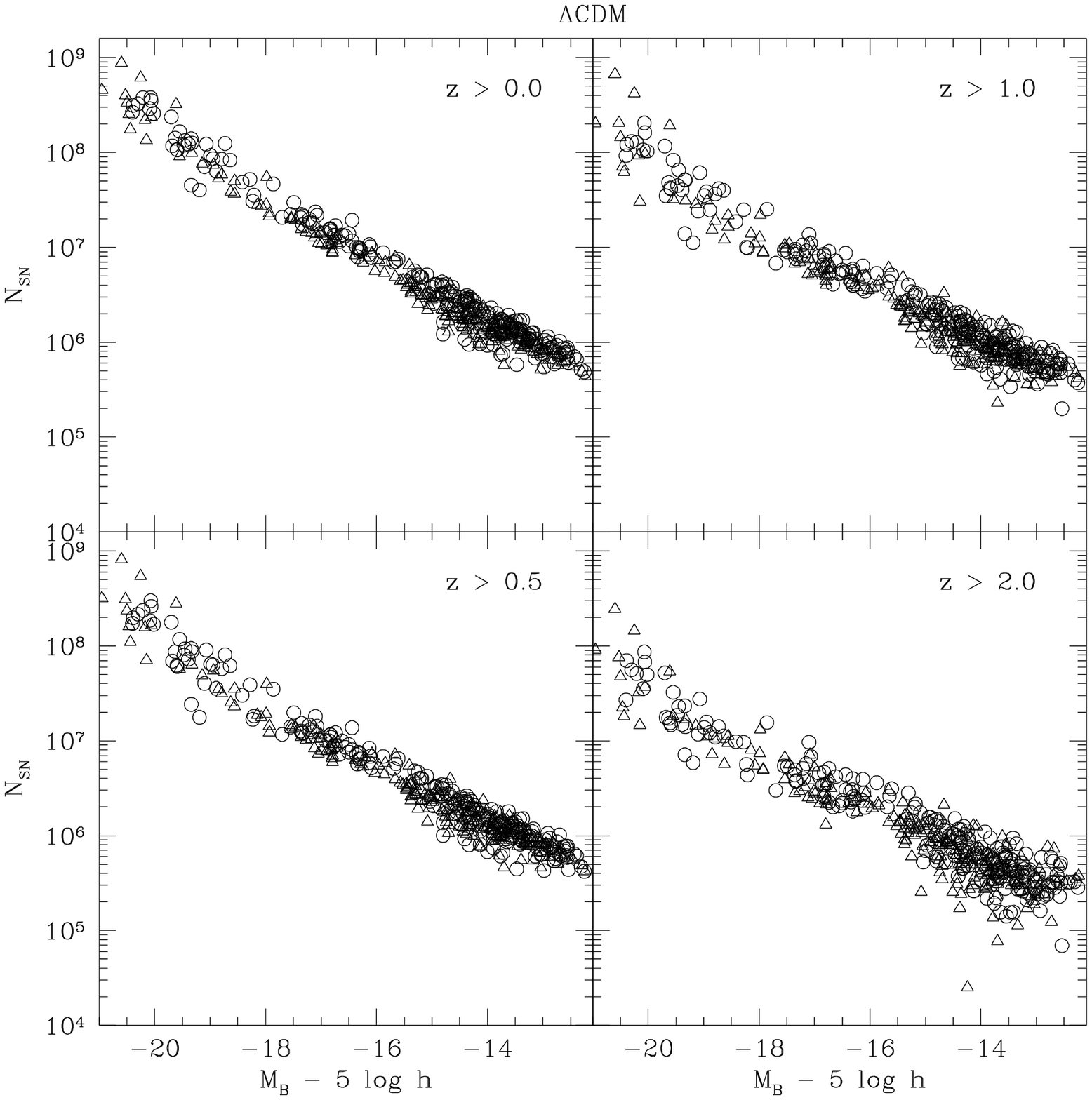}}
   \caption{Correlation of the $z=0$ $B$-band absolute magnitude of
     simulated galaxies and number of SNe type II exploded in the
     galaxy prior to a given epoch $z$ ($z=0$: upper left panel;
     $z=0.5$ lower left panel; $z=1$: upper right panel; $z=2$: lower
     right panel. Symbol labeling is the same as in Fig.~\ref{nsnm}.}  
    \label{nsnm_evol}
   \end{figure*}

The redshift dependence of the global star formation rate averaged
over all simulations is shown in Fig.~\ref{sfr} together with the
current data on observed global star formation history in the Universe
(see also \pcite{yepesetal99:sfr} for comparisons of other
cosmological models). The observational data were collected from
\scite{gallego95:sfr,lilly96:sfr,connolly97:sfr,sawicki97:sfr,hughes98:sfr,glazebrook99:sfr,madau98:sfr,pascarelle98:sfr,tresse98:sfr,treyer98:sfr,cowie99:sfr,steidel99:sfr},
and \scite{yan99:sfr}.  All data points correspond to measurements of
comoving UV or $H\alpha$ luminosity densities. In order to transform
to star formation densities, we have followed Madau's prescription
(\pcite{madau98:sfr}) to correct the original data for dust extinction
and to transform the luminosity densities to star formation densities.
All data points were properly rescaled to the $\Lambda$CDM
cosmological model used in the numerical simulations.  The figure
shows that the simulation results are in agreement (within the errors)
with the observed evolution of the global star formation rate.  The
star formation rate in the simulations may actually be a little higher
than the average observed rate, implying hence a larger number of
exploded SNe. As we derived the SN rate from the galactic 
star formation rate shown in Fig.~1, this figure may serve as an 
illustration of how the SN explosion rate evolves with time. 

We have analyzed two additional simulations to assess the effects of
resolution and box size. Particularly, the effects of resolution were
checked by re-running one of the $5$ Mpc simulations with $256^3$
grid cells and particles (i.e., with eight times better mass
resolution and twice the dynamic range). We have not found any
significant changes in the global star formation rate or in the
predicted number of supernovae (see below).  To test the effects of
the box size, we ran a simulation of $8.4\hMpc=12{\rm\ Mpc}$ box
using $300^3$ grid cells and particles, which gives the same
resolution as the $128^3$ $5$ Mpc runs but in a $2.4$ times larger
box. The results of this simulations are shown in Figs.~\ref{sfr} and
\ref{nsnm} together with the results of other runs. The figures show
that results of the large-box simulation are in agreement with results
of $5$ Mpc runs.

As we mentioned above, to estimate $\esng$ expected from galaxies
which end up in a cluster, we make use of the correlation between
absolute magnitude of a galaxy at z=0 and number of type II SNe
exploded in this galaxy throughout its evolution. The number of type
II SNe exploded in a galaxy of absolute magnitude $M$, $N_{SN}(M)$, is
computed as the fraction of gas mass converted into stars of mass
$\geq 10 M_{\odot}$ divided by the IMF-weighted mean SNe mass. We use
the \scite{salpeter55} IMF, for which these numbers are $0.12$ and $22
M_{\odot}$ (using lower and upper integration limits of $0.1
M_{\odot}$ and $125 M_{\odot}$), respectively. 

Figure~\ref{nsnm} shows $z=0$ correlation $N_{SN}(M_B)$ for galaxies
formed in the eleven $5$ Mpc and one $12$ Mpc {\LCDM} runs. The
correlation at $z=0$ can be well fitted by a linear fit (shown by
solid line) of the form $\log(N_{SN})=a+bM_B$, with $a=1.411$ and
$b=-0.344$.  Figure~\ref{nsnm_evol} shows evolution of this
correlation with redshift.  The figure shows that by $z=2$ the number
of exploded SNe is predicted to be $\approx 3-5$ smaller than the
number exploded by $z=0$. Note that galaxies in our simulations are
either isolated or are located in poor groups. It can be expected that
formation of cluster galaxies occurs somewhat earlier than that of
galaxies in poorer environments (by about $\Delta z\approx 1$, see,
e.g., \pcite{gkk:merging}) and the results for $z>2$ should probably
be interpreted as $z>3-3.5$ instead. 

To estimate the supernovae energy input in a cluster of a given virial mass, $\Mvir$, 
we convolve $N_{SN}(M_B)$ fit with the \scite{schechter76:lf} galaxy
luminosity function 
\begin{equation}
\phi(M)=0.4\ln 10\phi_{\ast}x^{1+\alpha}e^{-x}; {\ \ }x\equiv 10^{0.4(M_{\ast}-M)};
\end{equation}
where normalization parameter $\phi_{\ast}$ is assumed to be equal to
$\Delta_{vir}$ times its field value. The parameter $\Delta_{vir}$ is
the expected virial overdensity in a given cosmological model and is
$\approx 334$ for the {\LCDM} model adopted for our
estimate (e.g., \pcite{lahavetal91:lambda,ekeetal96:cluster}). The energy input is thus
\begin{equation}
\esng(\Mvir)=E_s\Delta_{vir}\left(\frac{4\pi}{3}\Rvir^3\right)
\int\limits^{M_b}_{M_f}N_{SN}(M)\phi(M)dM;
\end{equation}
where $E_s=\epsilon_{II}E_{SNII}$ is the energy input of a single supernova explosion, 
$\phi(M)$ is the field LF, $\Rvir$ is the virial radius of the cluster,
$M_b$ and $M_f$ are the  bright and faint limits of integration, and
$\epsilon_{II}$, $E_{SNII}$ have the same meaning as in the previous section.

The parameters of the luminosity function of galaxies in clusters
appear to be similar to those of the field LF (\pcite{trentham98:lf})
and we will therefore neglect possible small differences between
cluster and field LFs and cluster-to-cluster variations. We adopt
parameters $M_B=-19.5$ and $\alpha=-1.2$ of the Schechter luminosity
function consistent with recent measurements of $B$-band LF in the
field (\pcite{dacosta94:lf,zucca97:lf}) and in clusters
(\pcite{trentham98:lf}). The faint-end slope $\alpha$ is somewhat
steeper than in LFs from most of other field surveys (e.g., 
\pcite{loveday92:lf,marzke94:lf,lin96:lf}). However, the steep
value $\alpha=-1.2$ better matches the LF of cluster galaxies
and the faint end slope of the LF of the simulated galaxies 
(see \pcite{elizondoetal99:apj}). Therefore, we adopt this value in 
our analysis along with the normalization of the field LF
$\phi_{\ast}^{field}=0.02h^3 {\rm Mpc^{-3}}$ (\pcite{zucca97:lf}).
This value may be uncertain by a factor of two (see Table 1 in \pcite{zucca97:lf}). 
The $\esng$ estimate presented below is proportional to
$\phi_{\ast}$ and can be simply rescaled for other values. We use
the integration limits $M_B^b=-22$ and $M_B^f=-14$. The results are
insensitive to adopting a brighter $M_b$ or a fainter $M_f$. For
consistency, we use $\epsilon_{II}=0.1$ and $E_{SNII}=1.2\times
10^{51} {\rm ergs}$ adopted in the previous section.

\section{Results}
\label{sec:results}

\begin{figure*}
   \resizebox{\hsize}{!}{\includegraphics{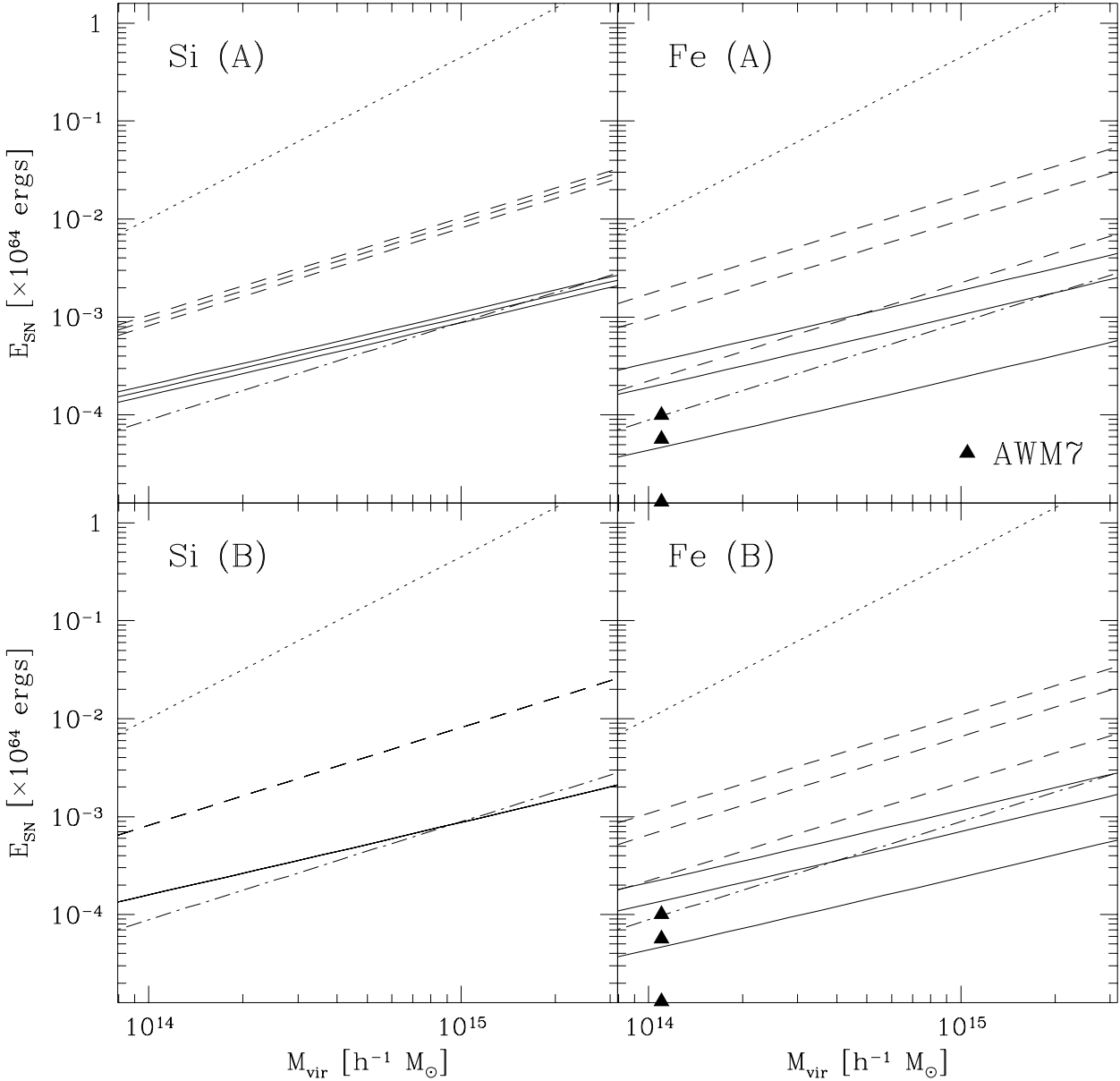}}
   \caption{Comparison of the energy input by SNe, $E_{SN}$, and
   thermal energy of the ICM gas, $E_{th}$, for clusters of different
   virial masses, $\Mvir$. The four panels show estimates of $E_{SN}$
   using observed metallicities of Si and Fe, as described in
   \S~\ref{sec:esnmetals}; {\em top row:} using yield model A, {\em
   bottom row:} using yield model B (see Table~\ref{table:yields}).
   $E_{SN}$ are shown for the cases where the ICM metallicity is
   uniform throughout the cluster ({\em dashed lines}) and for the
   presence of a strong metallicity gradient ({\em solid lines}) (see
   \S~\ref{sec:esnmetals} for details). {\em Solid triangles} are
   estimates for the cluster AWM7, in which such gradient was
   observed. $E_{SN}$ was estimated for three different metal
   fractions contributed by type Ia SNe: $f_{X_i}^I=0.0,0.5,1.0$,
   corresponding to the shown groups of three curves and points (the
   top curves/points correspond to $f_{X_i}^I=0.0$ and the bottom to
   $f_{X_i}^I=1.0$).  $E_{th}(\Mvir)$ ({\em dotted line}; the same in
   all panels) was calculated assuming the gas is distributed
   similarly to the dark matter (NFW distribution). $E_{th}$ of the
   gas distributed with the ``$\beta$-model'' is larger by $\sim
   10-20\%$. The energy released by type II SNe in the numerical simulations
   is shown by {\em dot-dashed lines} (the same in all panels).  }
   \label{esn} \end{figure*}

\begin{figure*}
   \resizebox{\hsize}{!}{\includegraphics{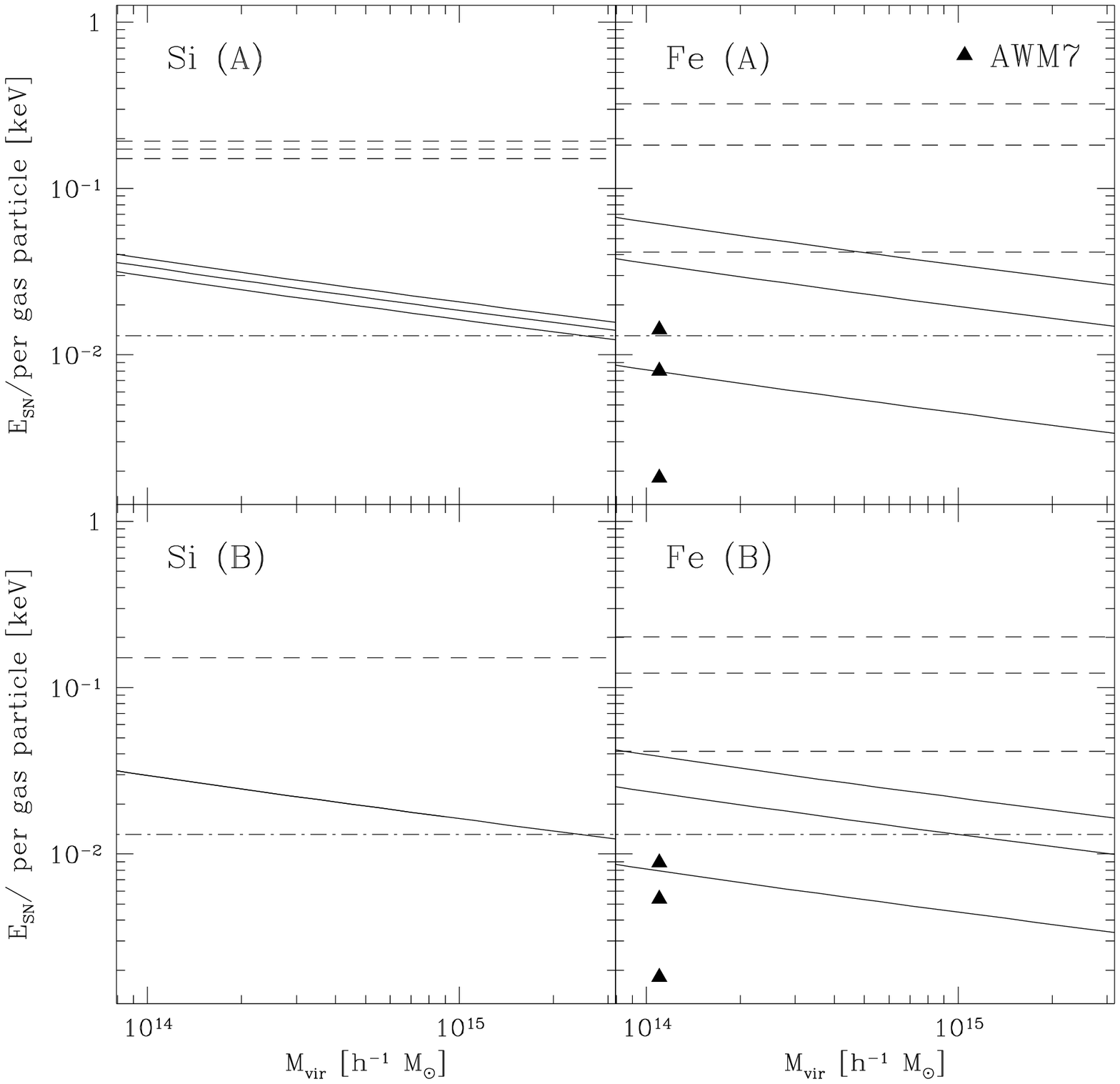}}
   \caption{Energy per gas particle injected by SNe ($=E_{SN}/N_p$,
   where $N_p=f_b\Mvir/(\mu m_p)$). The curves and points have the same 
   meaning as in Fig.~\ref{esn}. The upper row shows the estimate of $E_{SN}$
   from the observed metallicity assuming model A for SNII yields, while the
   bottom row shows the same estimate assuming model B.  }
   \label{esnpp} \end{figure*}

Figures~\ref{esn} and \ref{esnpp} show results of the estimates
described in the previous two sections. In Figure~\ref{esn} we compare
estimated energy input from SNe with the thermal energy of the ICM
gas. Top row of Fig.~\ref{esn} shows $E_{SN}$ estimate using observed
ICM metallicities and model A for SNII yields, while the bottom row
shows the same estimate for yield model B (see
Table~\ref{table:yields}).

The thermal energy of the gas is computed as
\begin{equation}
E_{th}=6\pi\frac{k}{\mu m_p}\int\limits_0^{\Rvir}\rho_g(r)T(r)r^2dr,
\end{equation}
where $k$ is the Boltzmann constant, $m_p$ is the mass of proton, 
$\mu=0.6$ is the assumed mean molecular weight of the ICM plasma,
$\rho_g(r)$ and $T(r)$ are its radial density and temperature
profiles. A density profile $\rho_g(r)$ is assumed, and the temperature
profile is calculated from the equation of hydrostatic equilibrium. In
Fig.~\ref{esn}, we assume that gas is distributed
similarly to dark matter, and is described by the \scite{nfw97}
(hereafter NFW), functional form,
\begin{equation}
\rho(r)=\frac{\rho_s }{(r/r_s)(1+r/r_s)^2}
\end{equation}
with appropriate scaling of parameter $c=\Rvir/r_s$ with cluster virial
mass (NFW). The observed distribution of the ICM gas is more often
described by the $\beta$-profile:
\begin{equation}
\rho(r)=\frac{\rho_0}{[1+(r/r_c)^2]^{3\beta/2}}, 
\end{equation}
where $r_c$ is the core radius, and the parameter $\beta$ controls the
outer slope of the distribution. The thermal energy of gas distributed
with the above density profile (for values of $r_c$ and $\beta$
consistent with the observed range) is only $\sim 10-20\%$ higher than 
$E_{th}$ of the NFW-distributed gas, the difference indistinguishable
on the scale of Fig.~\ref{esn}. 

Figure~\ref{esn}  shows that expected energy input from SNe is, depending
on the assumed uniformity of the ICM metallicity, $\sim 5-10\%$ of the
gas thermal energy for poor clusters ($\Mvir\sim 10^{14}\Msun$) and
$\lax 5\%$ for rich clusters ($\Mvir\sim 10^{15}\Msun$). In case of the
strong metallicity gradients these numbers are $\approx 3\%$ and $\lax
1\%$, respectively.  The estimates from both Si and Fe agree very well 
between each other, for $f_{Fe}^I\lesssim 0.5$.
Supernova energy input estimated from the numerical simulations 
agrees well with $E_{SN}$ derived from observed metallicities in
the case where a strong metallicity gradient is allowed. This
implies that if metallicity gradients exist in clusters, simple galaxy
formation models with the Salpeter IMF will have no difficulty in
accounting for the observed amount of metals in clusters. 

Figure~\ref{esn} also shows estimates for the cluster AWM7 for which a
large-scale abundance gradient has been observed and its parameters
measured (\pcite{ezawa97:awm7}).  The estimate was made using the
observed metallicity gradient: $Z(r)=Z_0[1+(r/r_c)^2]^{-3\beta_Z/2}$
with $Z_0=0.59$ and $\beta_Z=0.8$, where the core radius
$r_c=57.5\hkpc$ is equal to that of the gas distribution. The gas
distribution is described by the similar $\beta$-profile with
$\beta_g=0.58$ (see \pcite{ezawa97:awm7} for details). We have made
estimates for different gas fractions, $f_b$, within the cluster
virial radius but the results are only mildly sensitive to a
particular value of $f_b$; the estimates shown in figs.~\ref{esn} and
\ref{esnpp} were made assuming $f_b=0.2$. The energy input estimate
for AWM7 lies even lower than the solid lines at this mass because to
calculate the latter we have assumed a core radius of $r_c=100\hkpc$,
which results in a higher mass of metals and consequently larger
estimated $E_{SN}$.

\begin{table*}
\caption{Estimates of numbers of SNe and masses in Fe and Si for clusters of 
different masses (all masses are in $\hMsun$). }
\label{param}
 \begin{center}
 \begin{tabular}{|lrrr} \hline
 Cluster mass               & $10^{14}$           &  $5\times 10^{14}$   & $10^{15}$\\\hline
$N_{\rm SN}$ (simulations)            & $7.8\times 10^9$    &  $3.9\times 10^{10}$ & $7.8\times 10^{10}$\\
$N_{\rm SN}$ (for 1 keV per particle) & $4.9\times 10^{11}$ &  $2.5\times 10^{12}$ & $4.9\times 10^{12}$\\
$M_{\rm Fe}$ (observed, assuming $Z$-gradient)& $2.7\times 10^9$ &  $8.8\times 10^9$   & $1.5\times 10^{10}$\\
$M_{\rm Fe}$ (observed, uniform distribution) & $1.4\times 10^{10}$ &  $7.0\times 10^{10}$   & $1.4\times 10^{11}$\\
$M_{\rm Si}$ (observed, assuming $Z$-gradient)& $2.1\times 10^9$ &  $6.9\times 10^9$   & $1.2\times 10^{10}$\\
$M_{\rm Si}$ (observed, uniform distribution) & $1.1\times 10^{10}$ &  $5.5\times 10^{10}$   & $1.1\times 10^{11}$\\\hline
 \end{tabular}
 \end{center}
\label{table:numbers}
 \end{table*}

For reference, table~\ref{table:numbers} gives predictions for the
numbers of type II SNe in clusters of different masses based on our
model (line 1), as well as the total masses of Fe and Si inferred from
observations with assumptions of metallicity gradient and uniform metal
distribution (lines 3-6). The table also gives the number of SNe
required to deposit 1 keV per gas particle into the ICM (line 2),
estimated assuming that average supernova deposits $1.2\times 10^{50}$
ergs. The mass of metals predicted in our model can be easily obtained
by multiplying the number of SNe in line 1 of
table~\ref{table:numbers} by the corresponding mass-weighted yield
given in Table~\ref{table:yields}. Note, however, that the predicted
number refers to the type II SNe only, and the predicted mass of
metals is thus only due to SNIIe.

The question we would ultimately like to address is whether the energy input
from SNe can noticeably affect the thermal state of the ICM. To answer
this question, we need to know what energy input is needed to account
for the observed properties of the ICM. It is not completely clear what
energy is required. However, on theoretical grounds
(\pcite{kaiser91:preheat,evrardhenry91}) it is known that model
predictions are in better agreement with the data when gas is assumed
to be preheated (by some non-gravitational process) at an early
moment. The preheating results in gas evolution corresponding to a
higher adiabat, which affects the evolution of the accreted gas (in
particular, some of the accreted gas may avoid being strongly
shocked).

Numerical simulations (\pcite{metzler94,nfw95,mohr97:sizet,pen98:tf})
and semi-analytic models of cluster evolution
(\pcite{cavaliere97:lt,tozzinorman99,balogh99:preheat,valageassilk99:entropy,wu99:heating})
confirm that preheating results in cluster properties that are more in
accord with observations. For example, to simulate SNe heating
\scite{pen98:tf} preheats the gas in his gasdynamic simulations by
injecting 1 keV of energy per nucleon of gas, or, for plasma with
primordial composition, $\approx 0.5 {\rm keV}$ per gas particle.
\scite{cavaliere97:lt} assume in their model that SNe preheat the
intergalactic gas to temperatures of $\approx 0.5-0.7$ keV, which
corresponds to $(3/2)kT\approx 0.75-1.0 {\rm keV}$ per gas particle.
\scite{balogh99:preheat} and \scite{wu99:heating} argue based on their
semi-analytic calculations that the energy injection of $\sim 2-3$ keV
per particle is required to bring model predictions in accord with
observations.

We can compare our estimate of $E_{SN}$ to these numbers calculating
the energy per gas particle as $E_{SN}/N_p$, where $N_p=f_b\Mvir/(\mu
m_p)$ is the number of gas particles within the virial radius of
cluster. Figure~\ref{esnpp} shows $E_{SN}/N_p$ for $E_{SN}$ estimated
from Si and Fe and from galaxy formation simulations. The figure shows
that the maximum energy per particle of $\approx 0.1 {\rm keV}$ can be
injected by SNe if the ICM metallicity is homogeneous, while in the case
of a strong metallicity gradient the typical energy per gas particle is
only a few tens eV. In particular, the estimate of $E_{SN}$ for the cluster AWM7 is 
only $\sim 0.002-0.01$ keV per particle. The corresponding estimate from galaxy
formation simulations is $\sim 10^{-2} {\rm keV}$ per particle. These
numbers are $\sim 5-20$ times smaller than the typical energy
injection assumed in the cluster formation models quoted above.

\section{Discussion}
\label{sec:discussion}

The results presented in \S~\ref{sec:results} allow us to assess the
conditions required for the SNe energy input to be important in galaxy
clusters. The primary conditions that are implied by our estimate of
$\esn$ from the observed abundances of Si and Fe are (i) large-scale
uniformity of the metal abundances throughout the cluster volume and
(ii) near $100\%$ efficiency in transfer of the energy of SN explosion
to the thermal energy of the IGM gas. The latter assumption is rather
unlikely and the energies derived from the observed abundances should
therefore be considered as the {\em upper} limits on the amount of SN
energy that could have heated the IGM.

There are but a few theoretical predictions and observational data
concerning the degree of uniformity of the metal distribution in
clusters. Based on the numerical simulations that include galaxy
feedback and metal enrichment, \scite{metzler94} and \scite{metzler97}
predict that large-scale metallicity gradients should exist in
clusters. On the observational side, \scite{ezawa97:awm7} observed
such a gradient in cluster AWM7. More recently, 
\scite{finoguenov99} reported similar large-scale ($R\lax 0.5-1{\ \rm Mpc}$)
metallicity gradients detected using {\sl ASCA} observations for several other
clusters. It is not clear, however, whether such gradients are ubiquitous. 
Our estimates of energy input from observed metal abundances
differ by a factor of $\sim 5-10$ if we assume a uniform distribution
of metals versus metallicity gradients of the type observed in AWM7.
New, deep observations of ICM metallicity profiles are therefore
crucial to make this estimate much more reliable.  With the launch of
the {\sl Chandra}\/ X-ray satellite, such observations should become
available. Our estimate of the SN energy input for AWM7 is two orders
of magnitude lower than energy input which seem to be required to
sufficiently preheat the ICM gas.

Incidentally, the existence of large-scale abundance gradients in
clusters would solve the problem of the total iron mass in clusters.
\scite{david97} and \scite{gibson97} show that if the contribution of
type I SNe to the iron production in clusters is relatively small, the
total iron mass in the ICM is too large to be explained by type II SNe
produced with Salpeter IMF. \scite{brighenti99:comparison} argue,
however, that this solution is unattractive because it makes it
difficult to explain the metallicities and radial abundance gradients
in massive elliptical galaxies. It is clear from our analysis that the
existence of large-scale abundance gradients in the ICM can reduce the
estimate of the iron mass by up to an order of magnitude, thereby
eliminating the need for a large number of SNII and flatter IMF.

The predictions of the SNe energy input, $\esng$, of the numerical
simulations of galaxy formation presented in this paper, although
consistent with observed evolution of the global starfomation rate in
the Universe, are somewhat lower than the estimate from the metal
abundances, $\esn^m$.  The estimates $\esng$ and $\esn^m$ agree
reasonably well if a metallicity gradient is assumed and the contribution of
type Ia SNe to the iron enrichment is $> 50\%$. In particular, the
$\esng$ estimate is actually higher than estimate of $\esn^m$ for
AWM7.  However, the energy input in this case is of the order
of $10-50$ eV per gas particle, which is far short of the energy
injection typically assumed to bring theoretical models in accord with
observations: $\sim 0.5-2$ keV per particle. The estimate will still
be short by a factor of $\sim 5-10$ even if $100\%$ of the energy of every SN
explosion goes into heating the ICM gas. 

The above conclusions are for clusters of virial mass $\Mvir\grtsim
10^{14}\hMsun$.  Figure~\ref{esn} shows that the ratio of predicted
SNe energy input to the thermal energy of the ICM gas increases by
about an order of magnitude as the mass is decreased from
$10^{15}\hMsun$ to $10^{14}\hMsun$.  This trend means that the SNe
energy input may be much more important for clusters of mass
$\Mvir\lesssim 5\times10^{13}\hMsun$ than for more massive
clusters\footnote{Note that this conclusion depends on our assumption
  that total luminosity of stars in clusters is proportional to the
  cluster mass (See \S~\ref{sec:esnsimul}). Although this assumption
  is reasonable, there is evidence that mass-to-light ratio of
  clusters and groups is a function of system mass. The data indicates
  that mass-to-light ratio of galaxy groups is somewhat 
  smaller than that of clusters (e.g., \pcite{bahcall95:whereisdm}).
  In this case our conclusion would not be changed.}.  The mass
$5\times10^{13}\hMsun$ corresponds to the ICM temperature of $\approx
2$ keV (e.g., \pcite{ekeetal96:cluster}), while deviations from
non-similarity are observed in real clusters for temperatures of
$\lesssim 2$ keV (\pcite{ponman99:entropy,balogh99:preheat}).
Nevertheless, it appears that quantitatively our conclusions will
stand for poor clusters.  The entropy of the preheated gas required to
explain observations is $\sim 100{\ }{\rm keV{\ }cm^2}$
(\pcite{ponman99:entropy}) which corresponds to an energy of $\approx
1.5 (n_e/10^{-3}cm^{-3})^{2/3}$ keV per particle, where $n_e$ is
electron number density.  Thus, the energy injection into the gas in
cluster cores ($n_e\sim 10^{-3}{\rm\ cm^{-3}}$) is about $1.5$ keV per
particle. Semi-analytical calculations of \scite{balogh99:preheat} and
\scite{wu99:heating} show that the energy injection required to
explain the data may be even higher: $\sim 2-3$ keV per
particle\footnote{\scite{balogh99:preheat} give an estimate of the
  required entropy of preheated gas as $S=kT/(\mu m_H
  \rho^{2/3})\approx 3.7\times 10^{33}{\rm \ ergs\ g^{-5/3}\ cm^2}$.
  This corresponds to $E_p=1.5\mu m_H\rho^{2/3} S\approx 3.5$ keV per
  particle if we assume a typical density of gas in cluster cores
  ($\rho\approx 10^{-27}{\rm\ g\ cm^{-3}}$) and the above value of
  entropy.}.

Such energy input is marginally consistent with our $\esn^m$ estimate
in the case of uniform metallicities and $\epsilon\approx 1$. For the
case of metallicity gradients, $\esn^m$ is more than an order of
magnitude lower. The energy input, $\esng$, predicted from numerical
simulations is even lower and is $\approx 0.1$ keV per particle even
for $\epsilon=1$. Therefore, the conclusion we draw from this analysis
is that it is unlikely that the energy input from SNe is sufficient to
preheat the intracluster gas to the required entropy, unless all of
the explosion energy goes into heating of the gas {\em and} metal
abundances are uniform throughout the ICM. Moreover, in light of the
$\esng$ estimates, the SN energy input can only be important if
starformation rate in cluster environments is a factor of 10 higher
than the average cosmic rate.  Similar conclusions were reached by
\scite{balogh99:preheat}, \scite{valageassilk99:entropy}, and
\scite{wu99:heating}. Recently, \scite{loewenstein2000} have also used
observed abundance of Si in the ICM to estimate possible SNe heating
and found that the implied SNe energies would not be sufficient to
heat the entire cluster gas to the required levels (note that this
estimate was done assuming uniform distribution of Si and
$\epsilon_{SN}=1$). He pointed out, however, that SNe could still be
the source of heating if only the gas in cluster cores was heated.  In
this case, heating would have to occur after or during formation of a
cluster, not at early epochs as was assumed previously, but
sufficiently early enough to be consistent with lack of evolution of
metal abundances at lower redshifts ($z\lax 1$;
\pcite{mushotzky97:metalevol}). Details and quantitative predictions
of such a scenario are yet to be worked out.

It is obvious that there are a number of uncertainties in our
estimates of the SNe energy input. The estimates of $\esn^m$ made with
the assumption of uniform ICM metallicity are by a factor $\sim 3-10$
higher than the corresponding estimates in the case when a strong
metallicity gradient is assumed.  This uncertainty not only makes the
$\esn^m$ estimate uncertain, but also hinders comparisons of metal
abundances predicted by galaxy formation models with observations.
This will likely be resolved in the near future with the advent of
new, deep X-ray observations of clusters, but it is a major limitation
at present.  Currently, only one robust measurement of large-scale
metallicity gradient has been obtained (\pcite{ezawa97:awm7}). This
cluster, AWM7, confirms the existence of strong metallicity gradients
and the estimate of $\esn^m$ for this particular cluster supports our
conclusions. It is not clear, however, how universal such gradients
are in clusters.

Note also that our estimates are based on average abundances of Si and
Fe from a large sample of clusters. Abundances in individual clusters
may vary by a factor of $\grtsim 2-3$.  Thus, for example, abundances
of Si and Fe (in solar units) vary in the range $\sim 0.1-1$ and $\sim
0.15-0.45$, respectively
(\pcite{mushotzky96:abundances,fukazawa98:abundances}). The energy
estimates for individual clusters may therefore also vary by a
corresponding factor.

The theoretical yields of Si and Fe from type Ia and type II SNe used in
our analysis depend on specifics of the explosion model. The Si yields
from SNIa may be uncertain by a factor of two
(\pcite{nomoto97:yieldsSNIa,nagataki98}), while all models predict similar
(to $\sim 10\%$) yields of iron. The yields of SNII for Si and Fe vary
by $\sim 30-40\%$ between different models (e.g., \pcite{nagataki98}).
Yield models A and B used in our analysis approximately represent the
range of predictions and should therefore provide a fair estimate of
uncertainty. Our conclusions hold for both yield models. 

The fraction of supernova explosion energy that can be available for gas 
heating is also rather uncertain. \scite{larson74:sn} and \scite{babulrees92:dwarfe}
give analytical arguments that this fraction should be $\sim 0.1$. 
These arguments are supported by recent direct numerical simulations of 
\scite{thornton98} who studied radiative losses of a SN remnant (SNR) for a 
grid of densities and metallicities of the ambient gas. The arguments
and simulations, however, assume spherically symmetric evolution of
SNRs in ambient gas of uniform density. The efficiency may be higher
if the topology of ambient gas density is very assymetric and the gas
has been swept up and preheated by previous, recently exploded SNe
(\pcite{larson74:sn}). This, for example, may be the case during a
strong starburst (e.g., \pcite{tenorio88:araa}). The parameter
$\epsilon$ is thus likely to be environment dependent and the average
value would be determined by the relative number of SNe exploding
during periods of quiescent star formation vs. the number of SNe exploding
in starbursts. Regardless of the actual value, considerable radiation
losses are expected and therefore it seems very unlikely that the
efficiency is close to $100\%$ ($\epsilon=1$).

Beside the problem of heating efficiency, it is also not clear how the
heated interstellar gas and released SN energy is transferred to the
IGM (or ICM). Several transfer mechanisms have been suggested. Gas may
be blown away from galaxies by supernova-driven winds
(\pcite{mathews71:winds,yahil73:winds,larson74:sn}) which subsequently
shock the IGM gas.  Evidence for winds is indeed observed in some
starburst galaxies (e.g., \pcite{heckman90:superwinds}). However, only
a small fraction of gas is expected to be blown away by starbursts in
massive galaxies (e.g., \pcite{matteucci95:abundances}, 
\pcite{maclowferrara99:winds}) and therefore
ejected gas can only constitute a small fraction of ICM.  Clearly, the
same questions arise when we consider how energy released by SNe can
actually heat the IGM. If only a fraction of this energy is delivered
to IGM, this effectively means a smaller value of $\epsilon_{SN}$ and
strengthens our conclusions.

The gas can also be transferred to the ICM by ram pressure (e.g.,
\pcite{gunngott72:ram}) and tidal stripping. The efficiency of ram
pressure in clusters is not well known. Recent numerical simulations,
however, suggest that it may actually be rather low
(\pcite{abadi99:ram}). The tidal stripping is probably the most
efficient mechanism of delivering ISM gas to the intracluster medium,
especially for low surface brightness galaxies
(\pcite{moore99:stripping}). However, in this case the gas is
transferred to the ICM relatively late, after the epoch of cluster formation,
when a sufficiently deep potential well is formed. This is in conflict
with high metal abundances observed in high-redshift clusters 
(\pcite{mushotzky97:metalevol,hattori97:dark}).

Recently, \scite{gnedinostriker97:metals} suggested that
metal-enriched gas can be ejected at early epochs during galactic
mergers. This mechanism may transfer metal-rich hot interstellar gas
into IGM, where it can be further heated by shocks developed during a merger
or after an encounter between ejected material and the ambient IGM gas.  Despite
the abundance of possible processes, it is not clear which process (or
combination thereof) is responsible for the transfer of gas from galaxies
into the intergalactic medium. It is clear, however, that this question
needs to be clarified if SNe are to be considered a viable source of
IGM heating.

We have made a number of assumptions to estimate $\esng$ from the
galaxy formation simulations. Changing some of these assumptions can
change the energy input estimate. First of all, our assumption of
Salpeter IMF directly affects the number of SNe per given mass of
formed stars.  IMFs Flatter than a Salpeter result in a larger number
of supernovae and thus in a larger energy input for the same star
formation rate.  For instance, a 10\% flatter slope with respect to
Salpeter's results in a 50\% increase in the number of SNe, given the
same low-mass limit of the IMF.  Indeed, a flatter IMF has been
suggested as an explanation for the observed iron abundances in
clusters (e.g., \pcite{david97,gibson97}).  However, note that
\scite{brighenti99:comparison} argue that flatter IMF is not
consistent with the evolution of elliptical galaxies.  The number of SNe
depends also on the low-mass limit of the IMF, although in a less
sensitive manner. Thus, an increase of the lower-mass limit by a
factor of 2 (from 0.1 to 0.2 $M_\odot$) results in an increase factor
of 1.3 in the number of SNe.

To calculate the number of SNe exploded during a Hubble time in all
cluster galaxies we have assumed that the number density of galaxies in
clusters is equal $(1+\Delta_{vir})$ times its field value. This means
that clusters represent the same fluctuation in number of galaxies as
in their total mass. Although this is a reasonable assumption, we note
that in the {\LCDM} model (as well as in other low-matter density CDM
cosmologies) studied here, a certain amount of anti-bias ($b\sim
0.5$) is required for the model to be consistent with observed galaxy
clustering (\pcite{kph96:lcdm,jenkins98:evolution,kk99:bias}). This
anti-bias arises primarily in the densest regions of galaxy groups and
clusters (\pcite{kk99:bias}).  For an anti-bias of $b\approx 0.5$ the
number density of galaxies would be two times lower than assumed in our
analysis, which would reduce the estimated energy input by a factor of
two.

We neglected possible differences between the shape of the field and
cluster luminosity functions. These differences appear to be rather
small for the $B$-magnitude LF used here (\pcite{trentham98:lf}), and we
therefore think that the uncertainty associated with this assumption is
relatively small.

A more important assumption is that the global star formation
histories of field and cluster galaxies are similar. At present, there
is no convincing evidence otherwise. \scite{balogh99:differential},
for example, argue that star formation activity in cluster galaxies is
not very different from that in the field.  They argue, in fact, that
field galaxies may produce more stars (and more type II SNe) than
cluster galaxies in which the star formation is being gradually turned
off after their infall onto cluster. This is in fact consistent with
theoretical predictions of \scite{kobayashi99:snrate} who present
models for the evolution of the SN rate in clusters and the field.
They predict that the rate in clusters is higher than in the field
only at $z \grtsim 3.5$, while at lower redshifts it is actually lower
due to a decreased contribution from SNe in spiral galaxies. Their
predictions for the overall starformation rate in clusters are almost
an order of magnitude lower than the starformation rate in the
simulations presented here at $z\lesssim 3.5$ and are higher at higher
redshifts. It seems unlikely, however, that their prediction can
account for the required tenfold increase in number of SNe because
only a small fraction of SNe in cluster galaxies explode at $\grtsim
4$.  We therefore conclude that possible differences in starformation
histories between cluster and field galaxies are too small to change
our conclusions.

Nevertheless, it is known that rich clusters have properties different
than if they would have simply had been constructed from massive
ellipticals and small galaxy groups
(\pcite{david97,renzini97:iron,brighenti99:comparison}).  Ellipticals
and galaxy groups appear to have smaller gas fractions and lower metal
abundances than rich clusters do. In particular, the ratio of iron
mass in the ICM to the total blue luminosity of cluster galaxies is
consistently higher for clusters than for groups
(\pcite{renzini97:iron}). It appears also that parameters of the
models of elliptical galaxies that are tuned to produce the observed
metal abundances in the ICM are inconsistent with abundance
measurements in individual ellipticals, the problem which cannot be
solved by adjusting the SNIa contribution to the metal enrichment
(\pcite{brighenti99:comparison}). These problems may indicate that an
important component is missing in our understanding of the ICM enrichment
history and cluster evolution. However, our conclusions about the
importance of SNe energy input can only change if the star formation
rate in the volume from which the cluster forms is significantly higher at
all epochs than that star formation rate in the field.

\section{Conclusions}
\label{sec:conclusions}

We have presented estimates of the possible energy input by supernovae
into the intracluster medium. Although these estimates are prone to a
number of uncertainties, we have defined conditions which determine
whether SNe can be a significant source of ICM heating. The following
main conclusions can be drawn from our analysis.

The SNe energy input, $\esn^m$, estimated from observed ICM abundances
of Si and Fe is only significant ($\sim 1$ keV per particle) when we
assumed that the distribution of metals in the ICM is uniform (no
significant radial gradients) {\em and}\/ that $\sim 100$\% of
individual SN explosion energy goes into heating the ambient gas
followed by negligible cooling ($\epsilon \approx 1$) (see
\S~\ref{sec:esnmetals}).  If large-scale metallicity gradients are
assumed in clusters, the estimated energy input is $\sim 0.1-0.5$ keV
per particle for $\epsilon=1$ and, correspondingly, $0.01-0.08$ keV
per particle for a more realistic value of $\epsilon=0.1$.

As an example, we present estimates of the energy input for the cluster AWM7
for which the abundance gradient has been measured. We find that the observed
abundance of iron in this cluster implies a SNe energy input of $\lesssim 0.01$
and $\lesssim 0.1$ keV per particle for $\epsilon=0.1$ and $\epsilon=1$, 
respectively. 

The energy input, $\esng$, estimated using self-consistent
three-dimensional numerical simulations of galaxy formation which
include effects of shock heating, cooling, SN feedback, and multi-phase
model of ISM, are $\approx 0.01$ and $\approx 0.1$ keV per gas
particle for values of efficiency parameter $\epsilon=0.1$ and
$\epsilon=1$, respectively.  These values are somewhat lower than the
values of $\esn^m$ (but are in good agreement with estimates for the
AWM7). Nevertheless, the two estimates agree reasonably well if the
existence of large-scale abundance gradients is assumed in
clusters. We therefore emphasize the importance of new measurements of
large-scale metallicity gradients for testing the theoretical models.

Our estimates of the SN energy input in all cases, except the case of
uniform ICM abundances and $\epsilon=1$, fall short of the energy
injection of $\sim 0.5-3$ keV per particle required to bring
theoretical models of cluster formation in accord with
observations. This suggests that supernovae are unlikely to be the
only source of the IGM heating and should possibly be supplemented (or
substituted) by some other heating mechanism. Similar conclusions have
been reached in recent studies of \scite{balogh99:preheat},
\scite{valageassilk99:entropy}, and \scite{wu99:heating}. 
\scite{valageassilk99:entropy} propose radiation from quasars as an alternative
heating mechanism. This opens discussion of new possible processes for what 
appears to be a required high-redshift preheating of the intergalactic medium. 

\section*{Acknowledgements}

We would like to thank Anatoly Klypin for useful discussions and
comments.  A.V.K. was supported by NASA through Hubble Fellowship
grant HF-01121.01-99A from the Space Telescope Science Institute,
which is operated by the Association of Universities for Research in
Astronomy, Inc., under NASA contract NAS5-26555. GY acknowledges
support from S.E.U.I.D under project number PB96-0029.  The numerical
simulations used in this paper were run at the Centro Europeo de
Paralelismo de Barcelona (CEPBA).

\bibliography{../../biblio/data/sn,../../biblio/data/sams,../../biblio/data/galform,../../biblio/data/galaxylf,../../biblio/data/kravtsov,../../biblio/data/satellites,../../biblio/data/bibliography}{}

\end{document}